\documentclass[conference]{IEEEtran}
\IEEEoverridecommandlockouts
\usepackage{cite}
\usepackage[utf8x]{inputenc}
\usepackage{amsmath,graphicx}
\usepackage{cite}
\usepackage{amssymb,amsfonts}
\usepackage{algorithmic}
\usepackage{textcomp}
\usepackage{xcolor}
\usepackage{caption,subcaption}
\usepackage{tikz}
\usepackage{multirow}
\usepackage{hhline}
\usepackage{diagbox}
\usepackage{hyperref}
\usepackage{pgfplots}
\pgfplotsset{compat=1.17}
\usepackage{mathrsfs}
\usetikzlibrary{arrows}
\usetikzlibrary{positioning}
\usetikzlibrary{3d}
\hypersetup{
	colorlinks=true,
	linkcolor=blue,
	filecolor=magenta,      
	urlcolor=cyan,
	pdftitle={On the Computation of PSNR for a Set of Images or Video},
	pdfauthor={Onur Keleş, M. Akın Yılmaz, A. Murat Tekalp, Cansu Korkmaz, Zafer Doğan},
	pdfstartview=Fit,
	pdffitwindow=True,
}
\def\BibTeX{{\rm B\kern-.05em{\sc i\kern-.025em b}\kern-.08em
		T\kern-.1667em\lower.7ex\hbox{E}\kern-.125emX}}

\graphicspath{{figures/}}
\begin{document}
	
	\title{On the Computation of PSNR \\ for a Set of Images or Video}
	
	\author{\IEEEauthorblockN{Onur Keleş$^1$\thanks{$^1$ O.K. and C.K. are supported by AI Fellowships provided by the Koç University - Turkish Is Bank (KUIS) AI Center.}, M. Akın Yılmaz, A. Murat Tekalp$^2$\thanks{$^2$A.M.T. acknowledges support from TUBITAK 1001 project 217E033, TUBITAK 2247-A National Leader Researchers Award (No. 120C156), and Turkish Academy of Sciences~(TUBA).}, Cansu Korkmaz$^1$, Zafer Doğan$^3$\thanks{$^3$Z.D. acknowledges support by the TUBITAK 2232 International Fellowship for Outstanding Researchers Award (No. 118C337).}} \vspace{6pt}
		\IEEEauthorblockA{Department of Electrical \& Electronics Engineering, Koç University, 34450 İstanbul, Turkey \\
			okeles19@ku.edu.tr}
		
	}
	
	\maketitle	
	\begin{abstract}
		When comparing learned image/video restoration and compression methods, it is common to report peak-signal to noise ratio (PSNR) results. However, there does not exist a generally agreed upon practice to compute PSNR for sets of images or video. Some authors report average of individual image/frame PSNR, which is equivalent to computing a single PSNR from the geometric mean of individual image/frame mean-square error (MSE). Others compute a single PSNR from the arithmetic mean of frame MSEs for each video. Furthermore, some compute the MSE/PSNR of Y-channel only, while others compute MSE/PSNR for RGB channels. This paper investigates different approaches to computing PSNR for sets of images, single video, and sets of video and the relation between them. We show the difference between computing the PSNR based on arithmetic vs. geometric mean of MSE depends on the distribution of MSE over the set of images or video, and that this distribution is task-dependent. In particular, these two methods yield larger differences in restoration problems, where the MSE is exponentially distributed and smaller differences in compression problems, where the MSE distribution is narrower. We hope this paper will motivate the community to clearly describe how they compute reported PSNR values to enable consistent comparison.
	\end{abstract} 
	\vspace{-6pt}
	
	\section{Introduction}
	\label{sec:intro}
	
	The mean-squared error (MSE) is often used as a measure of fidelity for images and video. However, its value depends on the scale (dynamic range) of the input images/video. For example, the magnitude of the MSE is different if it is computed between the same two images, when they are normalized to $[0,1]$, or normalized to $[-1,1]$, or unnormalized $[0,255]$. Peak signal to noise ratio (PSNR) is used as a measure of fidelity that is independent of the dynamic range of images/video.
	
	In the learned image/video restoration and compression literature, the performance of methods is often reported by the arithmetic mean of the PSNR of each image/frame in the test set. For example, in the case of well-known EDSR image super-resolution model, the average PSNR for all images in the test set was reported~\cite{edsr2017}. In the case of EDVR video super-resolution model, the average of PSNR values for all frames in all videos were reported as a single value~\cite{edvr2019}. This is in contrast with the classical video compression literature, where a different PSNR is computed for each video using the average MSE of frames in that video. The difference between computing arithmetic mean of frame PSNR values versus computing PSNR based on arithmetic mean of frame MSE values for computation of PSNR for a single video  was investigated in \cite{nas2014} and it was concluded the latter measure correlates with human evaluation of video quality better.
	
	This paper discusses various nuances in the computation of PSNR for a set of images or set of videos in the context of evaluation of learned image processing methods, unlike~\cite{nas2014}, which only addressed computing PSNR for a single video in the context of video compression. In Section~\ref{sec:theory}, we present two methods to estimate PSNR for a set of images and the~relation/differences between them, which depend on the~distribution of MSE values over a set of images/video. We~discuss the same issues for a single video and a set of videos in Section~\ref{sec:video}. Section~\ref{sec:conc} concludes the paper.
	
	
	\section{Computation of PSNR for a Set of Images}
	\label{sec:theory}
	
	\subsection{Two Estimates}
	\label{sec:two estimates}
	
	Without loss of generality, let $X(n_1,n_2) : n_1, n_2 \in \mathbb{N}$ be 2-D discrete random field, that is an indexed set of random variables $x(n_1,n_2)$ representing pixel intensities of an image in the range $[0,1]$. Let 
	\begin{equation}
		\label{eq:mse}
		f_{X}(\hat{X})
		=\frac{1}{K} \sum_{(n_1,n_2)} \left( X(n_1,n_2)-\hat{X}(n_1,n_2) \right)^2 
	\end{equation}
	where $\hat{X}(n_1,n_2)$ is a reconstruction of $X(n_1,n_2)$, denote the~sum of square errors over all pixels~$(n_1,n_2)$; i.e., mean squared error. Clearly, $f_{X}(\hat{X})$ is a random variable since it is a function of a random field $X$.
	Let the random variable
	\begin{equation}
		\label{eq:psnr}
		g\left(f_{X}(\hat{X})\right) = -10 \log_{10}\left( f_{X}(\hat{X})\right) 
	\end{equation}
	denote the PSNR, where $\max \{\hat{X}(n_1,n_2) \} = 1$.
	
	We would like to obtain an estimate of the PSNR for an ensemble of images. It turns out there are two alternative ways to do this that give us different estimates: We can directly estimate $\mathbb{E}\{g(f_{X}(\hat{X}))\}$ or first estimate $\mathbb{E}\{f_{X}(\hat{X})\}$ and then compute $g(\mathbb{E}\{f_{X}(\hat{X})\})$. 
	
	Let $\operatorname{\overline{PSNR}}$ denote an estimator of $\mathbb{E}\{g(f(X))\}$ based on $N$ sample realizations (test images). Then,
	\begin{equation}
		\label{eq:psnr_pi}
		\mathbb{E}\left\lbrace g\left(f_{X}(\hat{X})\right)\right\rbrace \approx \operatorname{\overline{PSNR}} = -\dfrac{1}{N}\sum_{k=1}^{N}10\log_{10}\left(\operatorname{mse}_{k}\right)
	\end{equation}
	where $\operatorname{mse}_{k}$ denotes the MSE of the $k$'th sample realization. We can rewrite \eqref{eq:psnr_pi} as 
	\begin{eqnarray}
		\operatorname{\overline{PSNR}}  
		& = -\dfrac{1}{N}10\log_{10}\left(\displaystyle\prod_{k=1}^{N}\operatorname{mse}_{k}\right) \nonumber \\
		& = -10\log_{10}\left(\sqrt[N]{\displaystyle\prod_{k=1}^{N}\operatorname{mse}_{k}}\right).
		\label{eq:geometric}
	\end{eqnarray}
	Hence, computing the sample mean of individual image PSNR values is equivalent to using the geometric mean of the individual image MSE in the PSNR formula.
	
	Alternatively, we can define another estimate of the PSNR for a set of $N$ images based on using the arithmetic mean of the individual image MSE, $\operatorname{\overline{MSE}}$, given by
	\begin{equation} 
		\label{eq:psnr_sm}
		g\left(\mathbb{E}\{f_{X}(\hat{X})\}\right) \approx \operatorname{PSNR}_{\overline{\operatorname{MSE}}} 
		= -10\log_{10}\left(\dfrac{1}{N}\displaystyle\sum_{k=1}^{N}\operatorname{mse}_{k}\right)
	\end{equation}
	
	At this point, one may ask which is a better estimate for the PSNR of a set of images and what is the relation between them. While there is no clear answer to which one is better, it is straightforward to study the relation between them.

	\subsection{Relation Between the Estimates}
	\label{sec:diff}
	
	According to the Jensen's inequality, the expected value of a convex function of a random variable is greater than or equal to the convex function of the expected value of the random variable. Mathematically, given $f$ and $g$ are convex functions of a random variable $X$, we have
	\begin{equation}
		\label{eq:jensens}
		\mathbb{E}\{g(f_{X}(\hat{X}))\}\geq g(\mathbb{E}\{f_{X}(\hat{X})\}).
	\end{equation}
	Note that $f_{X}(\hat{X})$ is always positive. Hence, we can readily conclude that
	\begin{equation}
		\operatorname{\overline{PSNR}} \ge \operatorname{PSNR}_{\overline{\operatorname{MSE}}}.
	\end{equation}
	
	Furthermore, it is also possible to study the difference between the two estimates quantitatively when the number $N$ of images in the set is large. From (\ref{eq:geometric}) and~(\ref{eq:psnr_sm}), the difference between the two estimates is given by
	\begin{equation}
		\label{eq:psnr_diff}
		\operatorname{\overline{PSNR}} - \operatorname{PSNR}_{\overline{\operatorname{MSE}}} = 10\log_{10}\left(\dfrac{\dfrac{1}{N}\displaystyle\sum_{k=1}^{N}\operatorname{mse}_{k}}{\sqrt[N]{\displaystyle\prod_{k=1}^{N}\operatorname{mse}_{k}}}\right)
	\end{equation}
	which is proportional to the ratio of the arithmetic mean and geometric mean of individual image MSE, $\operatorname{mse}_{k}$.
	
	It is well-known that the geometric mean deviates from the~arithmetic mean as the variance of the numbers increases. It~was shown in \cite{Ogun2018} that in image restoration and super-resolution (SR) problems the variance of MSE values over a test set can be as high as the mean of the MSE values. The~dependence of the~difference between two PSNR values on the distribution of MSE is further analyzed  in the following.
	
	\subsection{Modeling the Distribution of MSE in a Set of Images}
	\label{sec:model}
	
	A properly-trained image restoration/SR model generates reconstructed images with low MSE for majority of the images in the test set. However, there are typically a number of images whose statistics differ from those in the training set resulting in higher MSE. Since the MSE is always positive and a large fraction of images have MSE close to zero, if the mean and the standard deviation of the MSE are close to each other, we can empirically model the distribution of MSE within the test set by an exponential distribution with parameter $\lambda$:
	\begin{equation}
		\label{eq:expdist}
		p(x) = \lambda e^{-\lambda x}\quad\text{for}\quad x\geq 0.
	\end{equation}
	Then, we can approximate the arithmetic and geometric mean of the MSE in \eqref{eq:psnr_diff} with those of exponential distribution.
	
	It is well-known that the mean of the exponential distribution is $1/\lambda$, the same as its standard deviation. According to \cite{vogel2020geometric}, its geometric mean is given by $e^{-\gamma}/\lambda$, where $\gamma$ is the~Euler-Mascheroni constant defined as
	\begin{align}
		\label{eq:euler_masch}
		\gamma &= \lim\limits_{n\to\infty}\left(-\ln(n) + \sum_{k=1}^{n}\dfrac{1}{k}\right) \approx 0.577216. \nonumber
	\end{align}
	
	Substituting these values into~\eqref{eq:psnr_diff}, we have
	\begin{eqnarray}
		\operatorname{\overline{PSNR}} - \operatorname{PSNR}_{\overline{\operatorname{MSE}}} 
		= & 10\log_{10}\left(\dfrac{1/\lambda}{e^{-\gamma}/\lambda}\right) \nonumber \\
		= & 10\log_{10}(e^{\gamma})\approx 2.506817.
		\label{limit}
	\end{eqnarray}
	
	This result shows that, independent of $\lambda$, generally reported sample mean of PSNR of individual images can be about 2.5 dB higher than the PSNR calculated from the arithmetic mean of individual image MSE, if the distribution of MSE is an exponential distribution. Since we approximate the distribution of MSE with a continuous probability density function, this value can be approached in the limit, where the number of images in the test set is large.
	
	
	\subsection{Examples}
	\label{sec:ex}
	
\begin{figure}[t]
		\centering
		\includegraphics[width=0.9\columnwidth]{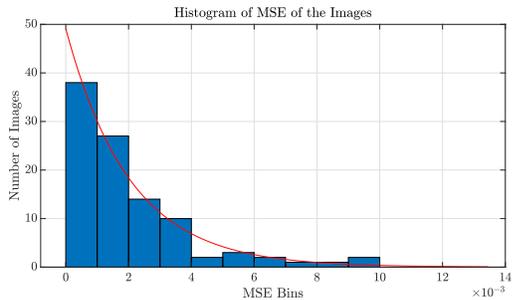} \vspace{-3pt}
		\caption{Histogram of the MSE values in the SISR task.}
		\label{fig:hist_mse}
	\end{figure}

	{\it Example 1: Single Image Super-resolution (SISR)}\\ We checked the hypothesis that the MSE is exponentially distributed over a set of test images in the case of an SISR task using the EDSR model~\cite{edsr2017} for $\times4$ SR on 100 validation images from the DIV2K dataset \cite{Agustsson_2017_CVPR_Workshops}. The~histogram of MSE values is shown in Figure~\ref{fig:hist_mse}, which resembles the exponential distribution. Moreover, the mean and standard deviation of the MSE for 100 images are $0.002037$ and $0.001978$, respectively, which support the assumption of  exponential distribution, since they are  close to each other. 
	
	\begin{table}[h]
		\centering
		\caption{Difference between $\operatorname{\overline{PSNR}}$ and $\operatorname{PSNR}_{\overline{\operatorname{MSE}}}$ for varying number of images in the case of the SISR task.}
		\begin{tabular}{|c||c|c|c|} 
			\hline
			No. of Images	& $\operatorname{\overline{PSNR}}$ & $\operatorname{PSNR}_{\overline{\operatorname{MSE}}}$ & Difference  \\\hline\hline
			30    & 28.9410 & 27.1707 & 1.7703 \\ \hline
			50    & 29.1912 & 27.1982 & 1.9930 \\ \hline
			100   & 28.9828 & 26.9109 & 2.0719 \\ \hline
		\end{tabular}
		\label{tbl:comparison}
	\end{table}
	Table \ref{tbl:comparison} shows that the difference between two PSNR values, computed in the RGB space, approaches the value predicted in~\eqref{limit} as the number of images gets larger.
	
	{\it Example 2: Image Compression}\\ The performance of image compression methods is evaluated by means of the rate-distortion function, where the rate and distortion are measured by the bits per pixel (bpp) and PSNR, respectively. Clearly, in this case, the distribution of the MSE values over a set of images depends on the value of bpp, and we expect to see the difference between the PSNR values calculated from the geometric and arithmetic means gets larger as the bpp gets smaller (QP gets bigger).
	This can be seen in Table~\ref{tbl:comparison_bpg}. Nevertheless, we observe that the difference between two PSNR estimates is lower in the compression task compared to the SISR task, since the distribution of MSE values is narrower even in the case of high QP.
	
	\begin{table}[h]
		\centering
		\caption{$\operatorname{\overline{PSNR}}$ and $\operatorname{PSNR}_{\overline{\operatorname{MSE}}}$ for BPG~\cite{bpg} compressed Kodak images.}
		\begin{tabular}{|c||c|c|c|} 
			\hline
			QP	& $\operatorname{\overline{PSNR}}$ & $\operatorname{PSNR}_{\overline{\operatorname{MSE}}}$ & Difference  \\\hline\hline
			30    & 36.22 & 36.07 & 0.15 \\ \hline
			35    & 32.88 & 32.61 & 0.27 \\ \hline
			40   & 29.81 & 29.43 & 0.38 \\ \hline
			45   & 27.14 & 26.68 & 0.46 \\ \hline
		\end{tabular}
		\label{tbl:comparison_bpg}
	\end{table}
	
	
	\subsection{Other Observations}
	\label{sec:observations}
	There are other factors that affect the PSNR values, including whether it is computed i) in the Y-domain or in the RGB domain, ii) in floating-point in the range $[0,1]$ or in uint8 in the range $[0,255]$, and iii) over a test set that contains images with mixed spatial resolution.
	
	The domain in which the PSNR is computed also varies in the community. Some SISR papers train models on RGB images but compute PSNR on Y channel only \cite{haris2019deep,luo2020unfolding,emad2021dualsr}, some both train models and evaluate PSNR on Y channel only \cite{cai2019toward}. Some others compute different metrics in different domains~\cite{wei2020component}, which makes fair comparison of methods difficult.
	
	While the authors in \cite{haris2019deep,luo2020unfolding,emad2021dualsr,cai2019toward,wei2020component} describe PSNR computations in detail, some SOTA papers on image compression \cite{cheng2020learned, balle2018variational, NEURIPS2018_53edebc5}, image denoising \cite{zamir2020cycleisp, yue2020dual, 8578952}, and image deblurring \cite{zamir2021multi}, do not explain how PSNR is calculated. Inspection of the provided source code reveals that in the image compression papers, images are first denormalized to $[0,255]$ and converted to uint8, then the PSNR is computed for each individual RGB image. On the other hand, in image denoising \cite{zamir2021multi}, PSNR for each reconstructed image is computed in floating point after clamping RGB images between $[0,1]$. Hence, the latter does not consider quantization error.
	
	Finally, some test datasets have images with mixed spatial resolutions. In that case, averaging PSNR values of images with mixed spatial resolutions may not be appropriate.

	
	\section{Computation of PSNR for a Set of Videos}
	\label{sec:video} 
	
	A video consists of a sequence of frames. Let's suppose we have a set of $N$ videos, each with $F$ frames in our test set. 
	
	One can treat each frame of each video as an independent image and report the PSNR of a set of $N \times F$ frames based on either approach discussed in Section~\ref{sec:theory}. Indeed, the common approach in the literature on learned video processing is to report the arithmetic mean of PSNR for all $N \times F$ frames \cite{edvr2019}.
	
	Alternatively, one can first compute an MSE or PSNR value for each video and then obtain a PSNR for the set of $N$ videos. We discuss these options in the following.
	
	\subsection{PSNR for Single Video}
	
\begin{figure}[h]
		\centering
		\includegraphics[width=\columnwidth]{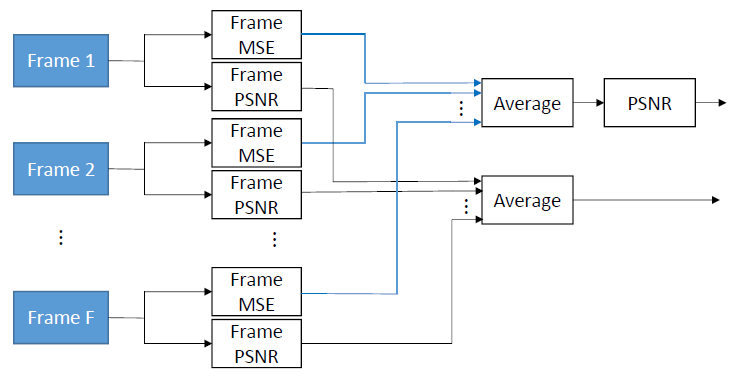} 
		\caption{Options for PSNR computation for single video.} \vspace{-4pt}
		\label{fig:videopsnr}
	\end{figure}

	Options for computing PSNR for a single video were discussed in \cite{nas2014}. They are i) by averaging individual frame PSNR values, or ii) computing PSNR based on the average MSE of all frames, which are summarized in Figure~\ref{fig:videopsnr}. 
	
	\subsection{PSNR for a Set of Videos}
\begin{figure}[t]
		\centering
		\includegraphics[width=\columnwidth]{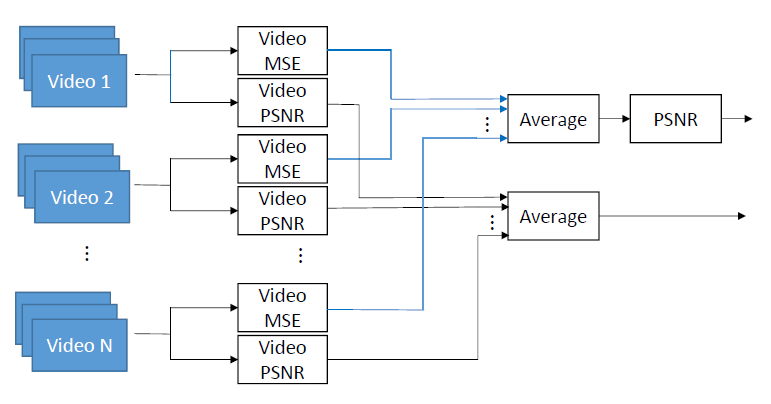} 
		\caption{Options for PSNR computation for a set of videos, where Video PSNR can be computed by either option in Fig.~\ref{fig:videopsnr}.}
		\label{fig:setvideo}
	\end{figure}

	Assume that we have a video test set with $ N $ videos, each of which has $ F $ frames. We can estimate the PSNR of the test set in three different ways for any video-related task:
	\begin{enumerate}
		\item \textit{PSNR-1}: Compute PSNR for each of $N\times F$ frames individually and then average them.
		\item \textit{PSNR-2}: Compute an MSE for each of $N$ videos, find the corresponding video PSNR value for each video, and then take the average of $N$ video PSNR values.
		\item \textit{PSNR-3}: Compute the average MSE for the test set from $N$ video MSE values, and then compute a single PSNR for the test set using the average MSE of the test set.
	\end{enumerate}
	Computation of PSNR-2 and PSNR-3 for a set of videos are illustrated in Figure~\ref{fig:setvideo}. Note that once we compute an MSE or PSNR for each video in the test dataset, the problem becomes the same as that addressed in Section~\ref{sec:theory}, and we use the two approaches discussed there to compute PSNR-2 and PSNR-3. 
	
	Most learned image processing methods use \textit{PSNR-1}, e.g., EDVR \cite{edvr2019}; whereas \textit{PSNR-2} is the method of choice adopted by standards-based video compression codecs. To the best of our  knowledge, \textit{PSNR-3} is not reported in any study before.
	
	\begin{table*}[t!]
		\centering
		\caption{Comparison of PSNR calculation for single video and the whole set from UVG dataset\cite{uvg} for video compression.}
		\resizebox{\textwidth}{!}{
			\begin{tabular}{|c|c||c|c|c||c|c|c||c|c|c||c|c|c|} 
				\cline{3-14}
				\multicolumn{2}{c|}{\multirow{2}{*}{}} & \multicolumn{3}{c||}{QP = 40} & \multicolumn{3}{c||}{QP = 35} & \multicolumn{3}{c||}{QP = 30} & \multicolumn{3}{c|}{QP = 25}  \\ \cline{3-14}
				\multicolumn{2}{c|}{} & \textit{PSNR-1} & \textit{PSNR-2} & \textit{PSNR-3}   & \textit{PSNR-1} & \textit{PSNR-2} & \textit{PSNR-3}   & \textit{PSNR-1} & \textit{PSNR-2} & \textit{PSNR-3}   & \textit{PSNR-1} & \textit{PSNR-2} & \textit{PSNR-3}    \\ \hhline{--*{12}{=}}
				\multirow{5}{*}{H264} & Video 1 \textit{Beauty}  & 30.60 & -- & 30.57 & 32.08 & -- & 32.07 & 33.06 & -- & 33.05 & 33.75 & -- & 33.73 \\ \cline{2-14}
				& Video 2 \textit{Jockey}  & 30.19 & -- & 29.98 & 32.88 & -- & 32.72 & 35.11 & -- & 34.99 & 36.82 & -- & 36.75 \\ \cline{2-14}
				& Video 3 \textit{Ready}  & 26.50 & -- & 26.34 & 29.24 & -- & 29.10 & 31.76 & -- & 31.64 & 34.16 & -- & 34.05 \\ \cline{2-14}
				& Video 4 \textit{Shake}  & 28.36 & -- & 28.30 & 30.20 & -- & 30.11 & 32.11 & -- & 32.01 & 34.00 & -- & 33.93 \\ \cline{2-14}
				& UVG Dataset & 29.08 & 28.94 & 28.64 & 31.35 & 31.19 & 30.99 & 33.41 & 33.26 & 33.13 & 35.28 & 35.14 & 35.02 \\ \hline\hline
				\multirow{5}{*}{H265} & Video 1 \textit{Beauty}  & 31.67 & -- & 31.66 & 32.74 & -- & 32.73 & 33.35 & -- & 33.34 & 33.99 & -- & 33.97 \\ \cline{2-14}
				& Video 2 \textit{Jockey}  & 32.74 & -- & 32.53 & 34.68 & -- & 34.55 & 36.33 & -- & 36.26 & 37.65 & -- & 37.62 \\ \cline{2-14}
				& Video 3 \textit{Ready}  & 28.30 & -- & 28.17 & 30.85 & -- & 30.73 & 33.28 & -- & 33.16 & 35.68 & -- & 35.57 \\ \cline{2-14}
				& Video 4 \textit{Shake}  & 29.02 & -- & 28.92 & 31.01 & -- & 30.89 & 32.97 & -- & 32.86 & 34.86 & -- & 34.79 \\ \cline{2-14}
				& UVG Dataset & 30.59 & 30.40 & 30.14 & 32.58 & 32.40 & 32.24 & 34.45 & 34.29 & 34.17 & 36.20 & 36.05 & 35.91 \\ \hline
			\end{tabular}
		}
		\label{tbl:videocomparison}
	\end{table*}
	
	\subsection{Examples}
	The common choice in the learned video SR community is to report PSNR-1 \cite{jo2018deep,edvr2019}. The authors in \cite{fuoli2019efficient} state that PSNR-2 would be a more appropriate choice but still report PSNR-1 in order to compare their results with others.
	
	In learned video compression almost all works report PSNR-1.
	Table \ref{tbl:videocomparison} shows comparison of three PSNR results for four individual videos as well as for the set of videos, which are encoded with two standard codecs. FFMPEG is used with \textit{veryfast} mode to compress each sequence and the settings in~\cite{dvc} are followed. Command lines for generating H264~\cite{h264} and H265~\cite{h265} compressed video are: \vspace{3pt} \newline
	\texttt{ffmpeg -y -pix}\_\texttt{fmt yuv420p -s 1920x1080 -i Video.yuv -c:v libx264 -preset veryfast -tune zerolatency -crf Q -g 12 -bf 2 -b strategy 0 -sc threshold 0 output.mkv} \vspace{3pt}  \newline
	\texttt{ffmpeg -pix}\_\texttt{fmt yuv420p -s 1920x1080 -i Video.yuv -c:v libx265 -preset veryfast -tune zerolatency -x265-params ”crf=Q:keyint=12” output.mkv} \vspace{3pt} \newline 
	respectively, where $Q$ denotes the QP value. As the Jensen's inequality predicts, the \textit{PSNR-1} gives the highest PSNR while \textit{PSNR-3} is the smallest. Similar to the image compression results, the differences between three PSNR values increase as the compression ratio increases. Note that \textit{PSNR-2} values for individual videos are not given since these results are the same with \textit{PSNR-3} for the individual videos.  In one of the early works, DVC~\cite{dvc}, the authors clearly state that they report \textit{PSNR-1} values. However, in more recent video compression works~\cite{hlvc,scale_space}, the details of the calculation of PSNR is not discussed. With the growing popularity of learned video compression, it is important that the authors clearly state the details of how PSNR is computed for a set of videos for fair comparison of different approaches.
	
	\begin{table}[h]
		\centering
		\caption{Comparison of PSNR calculation for frame prediction results for MPEG sequences presented in \cite{frame_pred}.}
		\begin{tabular}{|c||c|c|c|} 
			\hline
			Video	& \textit{PSNR-1} & \textit{PSNR-2} & \textit{PSNR-3}  \\\hline\hline
			Video 1: \textit{Coastguard}  & 31.99 & -- & 30.68 \\ \hline
			Video 2: \textit{Container}  & 41.50 & -- & 40.73 \\ \hline
			Video 3: \textit{Foreman}      & 31.38 & -- & 29.85 \\ \hline
			Video 4: \textit{Tennis}      & 29.88 & -- & 26.10 \\ \hline
			MPEG Dataset   & 32.88 & 30.03 & 29.08 \\ \hline
		\end{tabular}
		\label{tbl:frameprd}
	\end{table}
	
\begin{figure}[t]
		\centering
		\includegraphics[width=\columnwidth]{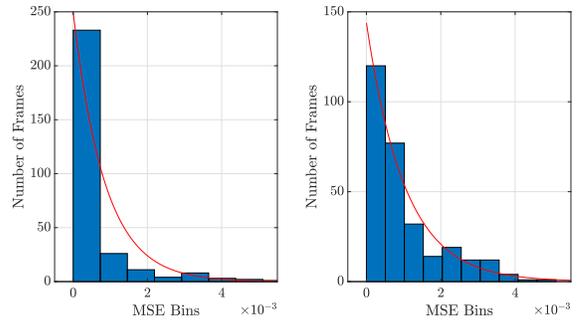}
		\caption{Histograms of MSE values for \textit{Coastguard} (left) and \textit{Foreman} (right) sequences for the next-frame prediction task.}
		\label{fig:cg_fm_hist}
	\end{figure}

	The differences between PSNR-1, PSNR-2 and PSNR-3 values in the case of learned next-frame prediction task are presented in Table~\ref{tbl:frameprd} for a test set with 8 MPEG videos. Fig.~\ref{fig:cg_fm_hist} depicts the histograms of frame MSEs for \textit{Coastguard} and \textit{Foreman}, which shows that they can be modeled by an exponential distribution. As a result, we see that the actual PSNR difference in the case of \textit{Foreman} video is close to the PSNR difference predicted by \eqref{limit}. This observation also holds for the PSNR difference for the complete test set.
	
	The results show that actual PSNR differences are close to predicted ones in both image and video SR and video next-frame prediction tasks, where the histograms of image/frame MSE values can be approximated by an exponential distribution. On the other hand, the distribution of image/frame MSE is narrower, and hence, the differences between PSNR values are smaller in image/video compression tasks.
	Hence, we conclude that distribution of image/frame MSE, and hence, the~difference between PSNR estimates is task dependent. 
	
	
	\section{Conclusions}
	\label{sec:conc}
	There are multiple ways to estimate the PSNR for a set of images, for a single video, and for a set of videos. We~recommend that the MSE~should be evaluated for samples of the random quantity defined; i.e., in the case of a set of images, MSE~should be computed per image sample and in the case of sets of video, MSE should be computed per video sample. Hence, PSNR for a single video must be computed from the average MSE for each video. Then, reporting PSNR for sets of images or video, whether computing PSNR from the average of MSE or average of individual PSNR values for each image/video as depicted in Figure~\ref{fig:videopsnr} and Figure~\ref{fig:setvideo}, respectively, is a choice. Hence, one needs to clearly specify how they compute reported PSNR values in comparing different learned image/video restoration/SR, and compression models. 
	
	We claim that in the case of $N$ videos with $F$ frames each, averaging frame-by-frame PSNR values for $N \times F$ frames, which is commonly done in the literature, is not a technically sound approach, since different video samples have different PSNR ranges depending on their motion features.
	
	Furthermore, in image restoration and SR tasks, the standard deviation of the MSE over a test set of images/video is close to the mean MSE; indicating an exponential-like distribution of MSE values, whereas in compression tasks we have a tighter distribution of MSE values. In any case, it is a good practice to report the variance of the MSE or PSNR values together with the mean MSE or PSNR values, as the mean MSE or PSNR alone may not be indicative of performance for all images/video in the dataset.
	
	\bibliographystyle{IEEEbib}
	\bibliography{refs}
	
\end{document}